\documentclass[11pt]{article}
\usepackage{moriond,epsfig}

\def\be{\begin{equation}}
\def\ee{\end{equation}}
\def\bea{\begin{eqnarray}}
\def\eea{\end{eqnarray}}

\def\ET{{\rm E}_{\scriptscriptstyle\rm T}}
\def\MET{\mbox{$\raisebox{.3ex}{$\not$}\ET$}}

\begin{document}
\vspace*{4cm}
\title{Searches for the Standard Model Higgs at the Tevatron}

\author{ B.Kilminster on behalf of the CDF and D\O \ Collaborations }

\address{Department of Physics, Ohio State University, 1040 Physics Research
Building, Columbus, Ohio 43210-1117}

\maketitle\abstracts{
The CDF and D\O \ experiments at the Tevatron are currently the only capable
of searching for the Standard Model Higgs boson.  This article describes
their most sensitive searches in the expected Higgs mass range, focusing on
advanced methods used to extract the maximal sensitivity from the data.  CDF
presents newly updated results for $H \to W^+W^-$ and $ZH \to l^+l^-
b\bar{b}$. D\O \ presents two new searches for $WH \to l\nu b\bar{b}$.  These
new analyses use the same 1 fb$^{-1}$ dataset as previous searches, but with
improved techniques resulting in markedly improved sensitivity.  }

\section{Introduction}

The Higgs boson is the last remaining Standard Model particle to be
discovered, and the one responsible for generating the $W$ and $Z$ gauge
boson masses.  Direct searches at the LEP experiments have excluded a Higgs
boson with mass less than 114.4 GeV/c$^2$ at 95\% Confidence Level (CL) in
the production mode $e^+e^- \to ZH$~\cite{Barate:2003sz}.  Experimental
measurements of the top quark and $W$ boson masses provide the strongest
indirect constraints on $m_H$.  Considering the newest CDF and D\O \ combined
top mass measurement of $m_t =$ 170.9 $\pm$ 1.8 GeV/$c^2$, and the newest CDF
$W$ mass measurement of $m_W =$ 80.398 $\pm$ 0.025 GeV$c^2$, in addition to
other precision electroweak other observables from LEP and SLD, the Higgs
boson mass is expected to be less than 144 GeV/c$^2$ at 95\% CL~\cite{eweak}.

The Tevatron at Fermilab provides 1.96 TeV center-of-mass energy from
proton-antiproton collisions in the two multi-purpose detectors, CDF and D\O
\ .  Gluon fusion is the highest cross-section process for producing a Higgs
boson, but because of high backgrounds at lower masses, this production
process is sensitive to Higgs mainly for $m_H >$ 135 GeV/$c^2$, where BR($H
\to W^+W^-$) starts at 68 \% and increases up to 90\% for $m_H =$ 160
GeV/$c^{2}$.  For $H \to W^+W^-$ in this mass range, the most sensitive final
state topology is two charged leptons with large missing transvere energy (
$\MET$).  For 114 $< m_H <$ 135 GeV/$c^2$, quark annhilation into an offshell
$W$ or $Z$ boson, which then emits a Higgs boson, provides the best
opportunity for discovery. At this mass range, the Higgs decays predominantly
$H \to b \bar{b}$. CDF and D\O \ have previously done 1 fb$^{-1}$ searches
for $H \to W^+W^- \to l^{+}\nu l^{-}\bar{\nu}$, $WH \to l\nu b\bar{b}$,
$ZH \to l^+l^-b\bar{b}$, and $ZH \to \nu\bar{\nu}b\bar{b}$, where $l= e,
\mu$. Because of the small expected Higgs signals, to maximize search
sensitivity over the allowed Higgs mass range, it is necessary to combine all
searches from both the CDF and D\O \ experiments, as well as improve analysis
techniques.

This proceeding outlines updates to searches in several of these channels,
focusing mainly on improvements in analysis techniques.  New CDF searches for
$ZH \to l^+l^-b\bar{b}$ and $H \to W^+W^- \to l^{+}\bar{\nu}l^{-}\nu$ are
presented for the first time, as well as two new D\O \ searches in the $WH
\to l\nu b\bar{b}$ channel.

\section{$H \to W^+W^- \to l^{+}\nu l^{-}\bar{\nu}$}
\label{sec:HWW}

\begin{figure}
\psfig{figure=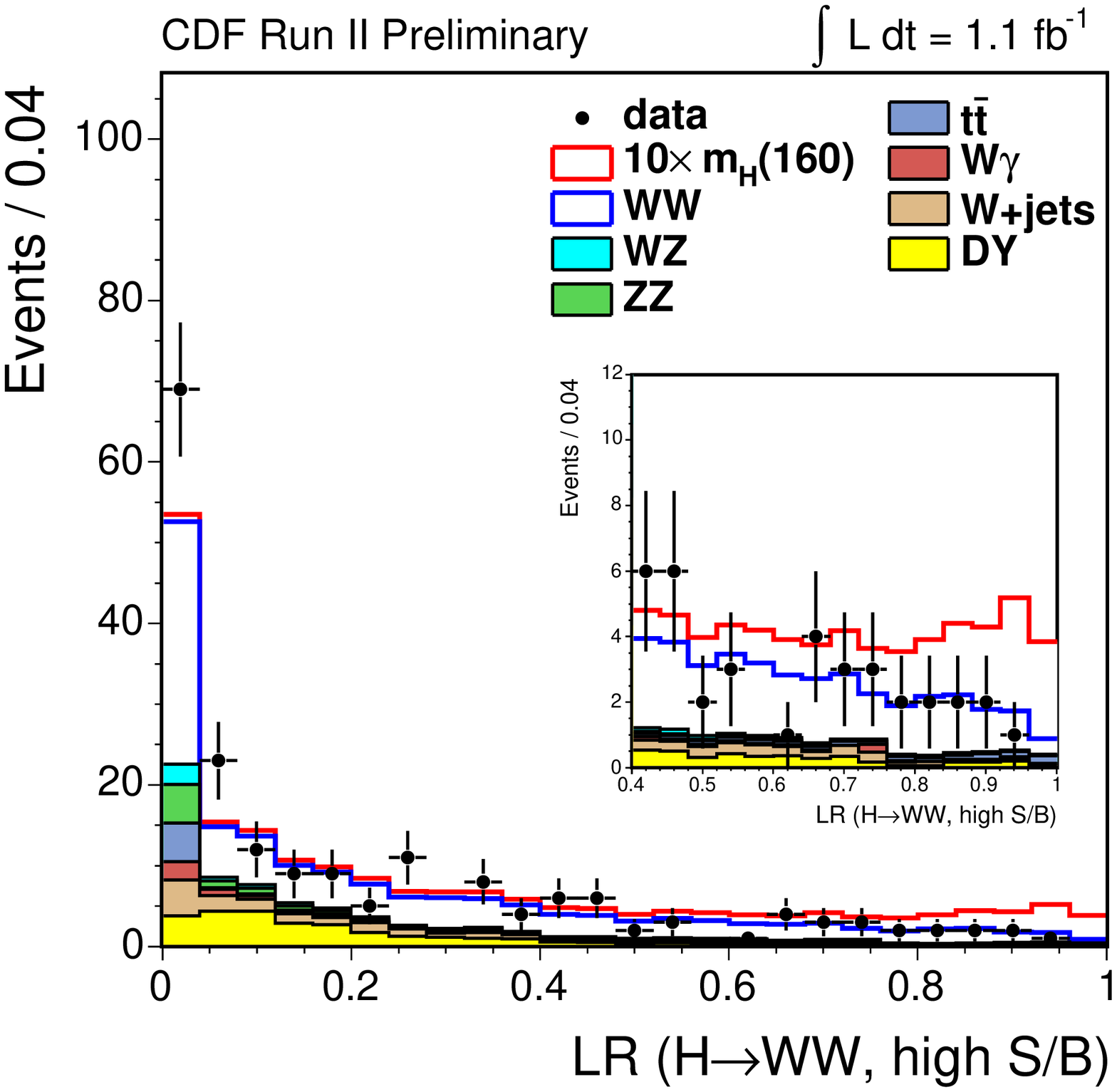,height=2.5in}
\psfig{figure=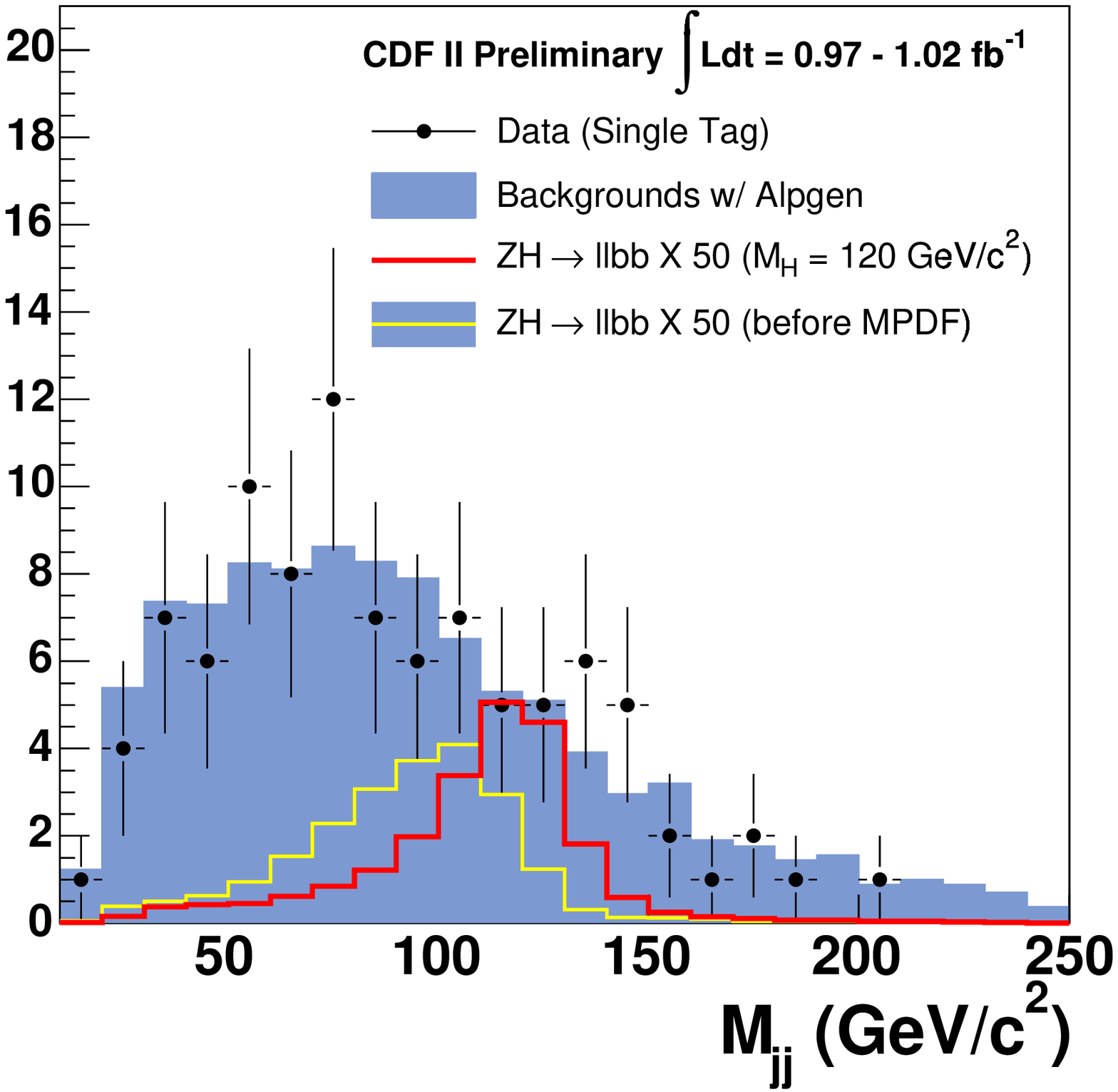,height=2.5in}
\caption{New CDF analyses. Left plot shows the $H \to WW$ discriminant, 
comparing data with Standard Model and with a 10 times expected $H \to WW$
signal.  Right, dijet mass improvement in the $ZH$ analysis shown before and
after the Met Projection Dijet Fitter (MPDF) is used to improve the
resolution of the dijet mass distribution, which is then input into the 2D
Neural Network discriminant used to fit the $ZH$ signal.
\label{fig:HWW}}
\end{figure}

CDF presents a new search in the $H \to W^+W^- \to l^{+}\nu l^{-}\bar{\nu}$
channel.  One improvement in this analysis is the increasing of geometric
lepton acceptance by defining new, less stringent lepton types for regions of
the detector without complete instrumentation, such as leptons not
identifiable as electrons or muons since they are not fiducial to
calorimeters or muon chambers. Such lepton types were successfully used in
CDFs observaton of $WZ$ production~\cite{Abulencia:2007tu}.  The number of
Higgs signal events expected in the data for $m_H$ = 160 Gev/c, increases
from 2.5 to 4.0 events with this new selection, as compared to the previous
CDF analysis with the same dataset.  The new analysis also improves upon the
technique for extracting the signal from the data. The previous search had
performed a likelihood fit for the Higgs signal using the distribution of
$\Delta \Phi$ between the two leptons which is sensitive to the angular
correlations from $WW$ produced by a scalar Higgs boson. The newest CDF
measurement is done by constructing matrix element probabilties using the
observed lepton four-vectors and $\MET$ for the processes $H \to WW$, $WW$,
$ZZ$, $W+\gamma$, and $W+$parton.  A likelihood ratio (LR) is formed for each
event by dividing the signal probability by the sum of the signal and
background probabilities.  LRs are constructed for different processes
specified as signal in order to validate background modeling. Figure
\ref{fig:HWW} shows the $H \to WW$ LR distribution for $m_H$ = 160 GeV/$c^2$,
which is used to search for an excess consistent with Higgs signal for a
range of masses.  No significant excess is measured, and limits are set such
that the observed (expected) upper limit is 5 (3.5) times larger than the
Standard Model expected cross-section for the most sensitive Higgs mass of
$m_H$ = 160 GeV/$c^2$~\cite{hww}.

\section{$ZH \to l^+l^-b\bar{b}$}
\label{sec:zhllbb}

CDF also presents a new result in the $ZH \to l^+l^-b\bar{b}$ channel.
Previous results first presented at ICHEP 2006~\cite{zhllbb-ichep}
demonstrated the use of a two-dimensional neural network trained to separate
$ZH \to l^+l^-b\bar{b}$ from the dominant background of $Z + >=$ 2 jets
production and $t\bar{t} \to WbWb \to l{\nu}b l{\nu}b$.  The new result uses
the same dataset but makes several improvements which result in improved
sensitivity.  One technique is in the identification of $b$ hadrons from $H
\to b\bar{b}$ using secondary vertex finders or ``$b$-tagging'' algorthms.
The previous $ZH$ analysis selected events with $>= 1$ ``$b$-tag'', using
tight requirements for the secondary vertex significance (40\% efficient).
Since S:B is 200:1 for events with one $b$-tag (40\% efficient), but 50:1 for
events with two looser $b$-tags (each 50\% efficient), there is an
improvement in Higgs sensitivity by fitting these classes of events
separately.  Another new technique is to improve the resolution of the $H \to
b\bar{b}$ dijet mass distribution, which is one of the most important Neural
Network inputs. Since the main cause of $\MET$ in $ZH \to l^+l^-b\bar{b}$
events is from jet energy mismeasurement, a correction is applied which
corrects the leading two jets independently according to their projection
onto the $\MET$ direction. The effect is to reduce the dijet mass resolution
from 14\% to 9\% for events with two $b$-tags.  With these two enhancements,
the analysis improves its Higgs search sensitivity by a factor of two in
terms of an effective luminosity increase as compared to the previous version
of the analysis with the same dataset. For $m_H =$ 115 GeV/$c^2$, the
observed (expected) upper limits for $\sigma_{ZH}$ are 16 times that of the
Standard Model~\cite{zhllbb-new}.

\section{$WH \to l \nu b\bar{b}$}

\begin{figure}
\psfig{figure=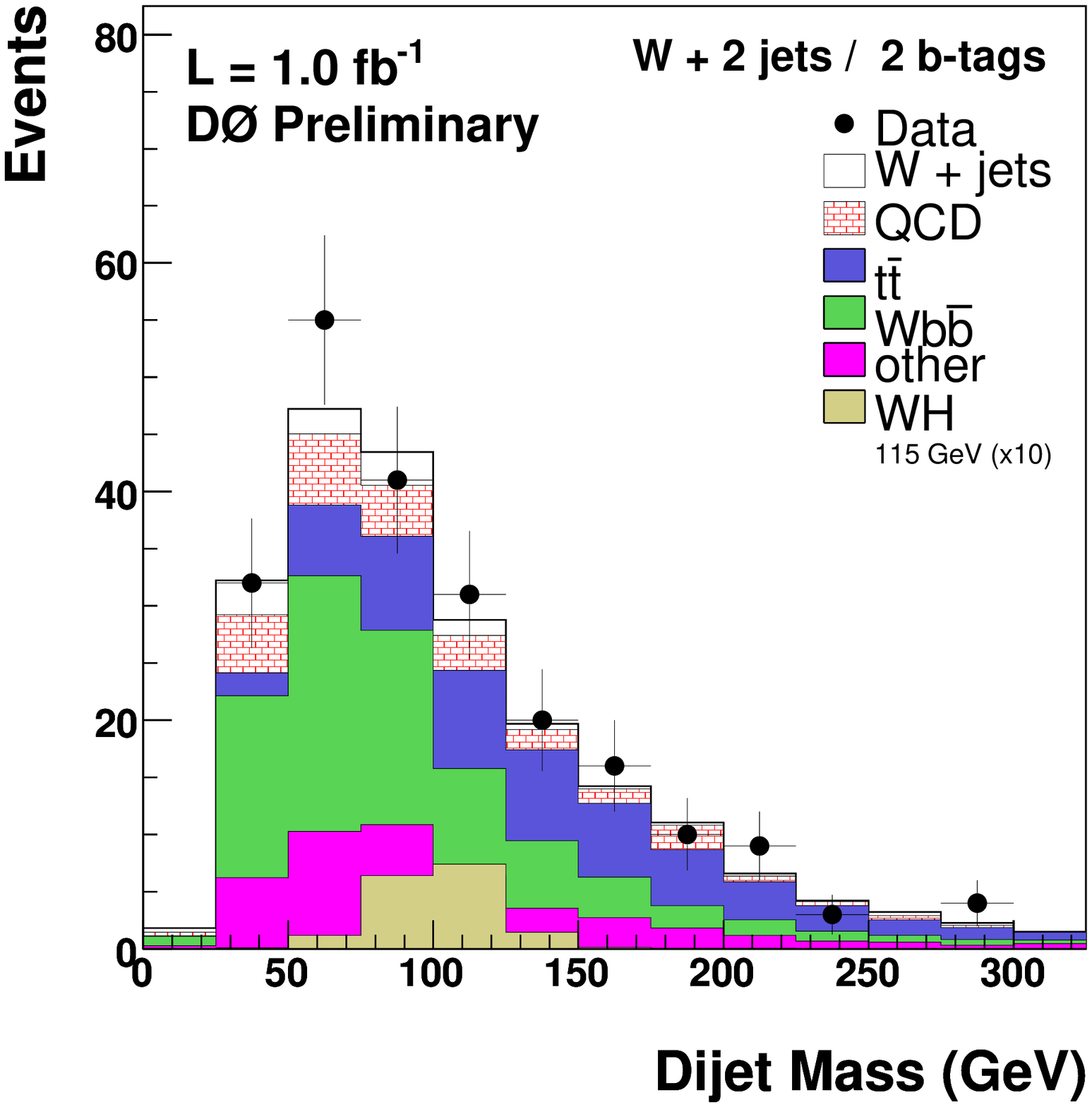,height=2.5in}
\psfig{figure=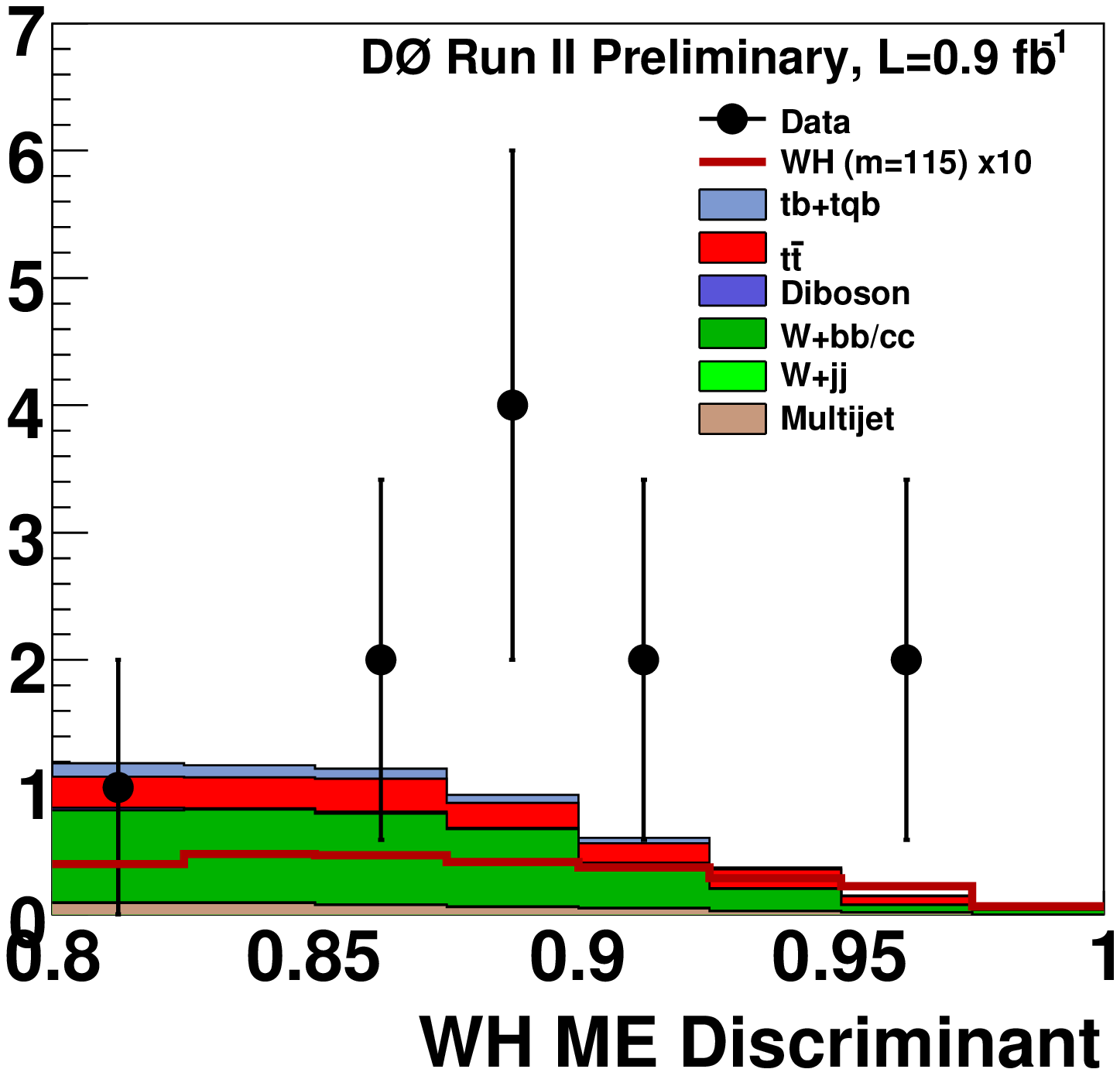,height=2.5in}
\caption{ New D\O \ analyses. Left, dijet mass distribution for $WH$ dijet
mass fit analysis, and right, discriminant for $WH$ matrix element 
analysis.
\label{fig:wd-d0}}
\end{figure}

D\O \ presents two new searches in the $WH \to l \nu b\bar{b}$ channels.  The
first analysis improves over previous analyses by using multiple muon
triggers in order to retain 100\% muon acceptance for $WH \to \mu \nu
b\bar{b}$.  This results in 50\% more signal than previous techniques. The
$b$-tagging selection is optimized to separate events into one tight $b$-tag
and two loose $b$-tags as is described in Section~\ref{sec:zhllbb}. But in
addition, making use of a neural network $b$-tagging algorithm which uses
variables in addition to the secondary vertex displacement to identify
$b$-quarks, the $b$-tagging efficiency is increased to 50\% for tight
$b$-tags and 70\% for loose $b$-tags, with misidentification rates of 0.5\%
and 4.5\%, respectively.  The discriminant used to search for the Higgs
signal is the dijet invariant mass (Figure~\ref{fig:wd-d0}), and the combined
limit from both single and double $b$-tagged events is expected (observed) to
be less than 9 (10) times the Standard Model expectation~\cite{wh-d0-jj}.

The second analysis makes use of a matrix element technique similar to the
one described in Section~\ref{sec:HWW}.  This matrix element technique was
developed originally in the context of the D\O \ single-top analyses which
established 3-$\sigma$ evidence for single-top production
\cite{Abazov:2006gd}.  By fitting the matrix element discriminant for $WH$
(Figure~\ref{fig:wd-d0}), D\O \ obtains expected (observed) limits of 9 (13)
times the Standard Model expectation~\cite{wh-d0-me}.  However, this analysis
does not make use of the improved muon acceptance and optimized $b$-tagging
used in the first analysis.  Incorporating these improvements into the matrix
element approach is expected to yield 30\% better expected sensitivity.

\section{Conclusions}

CDF and D\O \ are improving analysis techniques in order to make gains in
Higgs sensitivity which scale much faster than increasing statistics alone.  New
results in $H \to WW$, $ZH$, and $WH$ rely on improved lepton acceptance and
triggers, higher efficiency $b$-taggers, and dijet mass resolution
improvements.  The new $WH$ and $H \to WW$ results are better in their most
sensitive mass ranges than the combined results of all CDF and D\O \ channels
presented a year ago.  

\section*{Acknowledgments}
We thank the Fermilab staff and the technical staffs of the participating
institutions for their vital contributions. This work was supported by the
U.S. Department of Energy and National Science Foundation; the Italian
Istituto Nazionale di Fisica Nucleare; the Ministry of Education, Culture,
Sports, Science and Technology of Japan; the Natural Sciences and Engineering
Research Council of Canada; the National Science Council of the Republic of
China; the Swiss National Science Foundation; the A.P. Sloan Foundation; the
Bundesministerium f\"ur Bildung und Forschung, Germany; the Korean Science
and Engineering Foundation and the Korean Research Foundation; the Particle
Physics and Astronomy Research Council and the Royal Society, UK; the
Institut National de Physique Nucleaire et Physique des Particules/CNRS; the
Russian Foundation for Basic Research; the Comisi\'on Interministerial de
Ciencia y Tecnolog\'{\i}a, Spain; the European Community's Human Potential
Programme under contract HPRN-CT-2002-00292; and the Academy of Finland.


\section*{References}
\begin{thebibliography}{99}

\bibitem{Barate:2003sz}
  R.~Barate {\it et al.}  [LEP Working Group for Higgs boson searches],
  Phys.\ Lett.\ B {\bf 565}, 61 (2003)
  [arXiv:hep-ex/0306033].

\bibitem{eweak} LEP Electronweak Working Group,
    http://lepewwg.web.cern.ch/LEPEWWG, March 2007 results.
\bibitem{Abulencia:2007tu}
  A.~Abulencia {\it et al.}  [CDF Collaboration],
  arXiv:hep-ex/0702027.
\bibitem{hww} CDF Collaboration. CDF public note 8774, 2007.
\bibitem{zhllbb-ichep} B.Kilminster [CDF and D\O \ collaborations], ICHEP 2006
proceedings, hep-ex arXiv:hep-ex/0611001.
\bibitem{zhllbb-new} CDF Collaboration. CDF public note 8742, 2007.
\bibitem{wh-d0-jj} D0 Collaboration. D0 public note 5350, 2007.
\bibitem{Abazov:2006gd}
  V.~M.~Abazov {\it et al.}  [D0 Collaboration],
  Phys.\ Rev.\ Lett.\  {\bf 98}, 181802 (2007)
  [arXiv:hep-ex/0612052].
\bibitem{wh-d0-me} D0 Collaboration. D0 public note 5365, 2007.



\end{thebibliography}
\end{document}